# The Causal Event Set

Alasdair Macleod[†]

ABSTRACT

To clarify some aspects of the application of Special Relativity, spacetime is sliced into null geodesic hypersurfaces as an alternative to the hypersurfaces of simultaneity normally adopted. Events at particle locations on the hypersurface are identified as the causal event set. It is demonstrated that a Lorentz boost applied to the causal event set maintains the property of connectedness and with this formalism it is simple to derive the redshift equation. The twin paradox is naturally explained as an instantaneous reconfiguration of particle position 4-vectors in the frame of the accelerated object. The metaphysical implications are examined with the tentative conclusion that a relationist view of spacetime is more consistent with this treatment than the substantivalist viewpoint.

[†] University of the Highlands and Islands, Lews Castle College, Stornoway, Isle of Lewis, Scotland, UK. (Alasdair.Macleod@lews.uhi.ac.uk).

**1. Introduction**

The Special Theory of Relativity unifies space and time and although the application of the basic postulates results in counter-intuitive predictions, these have been fully verified by experiment [1]. However, in spite of the overwhelming evidence, there are still some who do not accept the fundamental principles of relativity (although the form of the criticism is often manifest through a disagreement with the process by which the relativistic transformation rules are derived). It is interesting to consider why this is the case. The theory is clearly unpalatable to many because it goes against common sense and there is certainly a lack of an ontological structure from which the principles of relativity naturally emerge. There is currently no answer to the question 'Why is the speed of light a constant?' though one would suspect that the question is key to a deep understanding of the Universe. It is no exaggeration to state that Special Relativity exists in a metaphysical vacuum. Metaphysics has offered the physicist little assistance in developing a rationale for the frame independence of the speed of light, or even why the laws of physics should be the same in all inertial frames.

In this paper some aspects of the criticism of Special Relativity will be reviewed and it will be shown that by changing the way spacetime is sliced, from the hypersurface of simultaneity to a null geodesic hypersurface, the application of relativistic transformations becomes practically easier and a metaphysical basis for the postulates begins to emerge.

**2. Paradoxes and Problems**

The two basic postulates of Special Relativity are that the laws of physics are the same in any inertial frame and that the speed of light is a constant independent of reference frame [2].

Whilst there is some debate as to whether space is a substantive medium in which particles are embedded or an artefact arising from the relationship between particles[¶], the prevalent viewpoint is that spacetime is substantial. Support for this comes from the theories of General Relativity (GR) and Quantum Mechanics (QM), the two most successful theories devised. Both require a 'real' space-time. In GR, space-time is assigned a local curvature, which elegantly explains gravitational effects. QM requires that empty space have zero-point energy if only to initiate 'spontaneous' emission.

In spite of the undoubted correctness of GR and QM, some situations are still analysed classically, using, for example, Maxwell's field equations. Here too a substantial space is required: action-at-a-distance is expressly forbidden; interaction is expected to satisfy 'spatiotemporal locality'[3] with the space between distant interacting entities occupied by fields to maintain spatiotemporal continuity.

The Special Theory of Relativity has revolutionised our understanding of space and time and shown that it is the space-time continuum that must be considered real – not just space itself. Separating space from time and assigning it form simply reintroduces the æther, a fluid with mechanistic properties that was proven not to exist by the Michelson-Morley and other experiments [§]. This distinction is crucial if the speed of light is to be independent of inertial frame and a failure to appreciate the difference is the source of much erroneous criticism.

Criticism of Special Relativity at a metaphysical level is actually rare. The critics instead tend to concentrate on particular problems and the mechanics of how information is acquired and conveyed. The focus is frequently on the so-called twin paradox (also known as the clock paradox). One form of the supposed paradox concerns two hypothetical twins A and B. Twin B is accelerated and travels to a distant star then decelerates and returns to Earth. Special Relativity predicts that twin B will have aged less than the stay-at-home twin A. The 'paradox' arises because from the stationary or rest frame of twin B, it is the Earth that appears to be moving, first away and then towards, with the result that twin B will predict that the twin on Earth will have aged less using the same argument as the sibling. This is obviously contradictory: they cannot both have aged less.

Clearly the situation is asymmetrical because of the acceleration of twin B, but the persistence of the paradox is astonishing - as recent as 2001, Dolby and Gull felt the need to explain away the paradox [5]. One reason for its persistence is that, having accepted that the acceleration on one twin is the cause of the asymmetry, it is not made clear by many authors how precisely it affects the calculations, particularly when it is well known that Special Relativity can be applied in a piecewise manner to situations involving acceleration without the need for corrections (even with accelerations as large as $10^{19}$ g, as demonstrated by Bailey *et al* [6]). A typical example of the lack of detail offered by some sources is the conclusion to Sciama's explanation of the paradox [7]:

> ".. the difference .. is that B has accelerated relative to distant matter while our stay-at-home A has not. …These considerations dispose of the clock 'paradox'. "

In 1971, Marder published a entertaining book [8] on the subject showing the resoluteness of some objectors of the time, particularly Professor Dingle. In the book, the resolution of the paradox is clear, but apparently not sufficiently to sway the critics.

There is no doubting the correct application of Special Relativity is difficult; even experienced scientists err. For example, in 1977, Davies and Jennison [9] measured the redshift of light reflected from a transversely moving mirror. A null result was obtained, in contradiction with some of the theoretical predictions from earlier literature. The authors noted that several theorists had incorrectly applied Special Relativity to the situation.

---

[¶] In ontological terms, the substantivalist viewpoint states that space (time) is an entity whilst the relationist view is that space is an expression of the relationship between entities. The differing viewpoints can be illustrated by considering the series of integers 1, 2, 3, .. and asking what lies between them. One substantivalist response (not the only one) might be to describe the fractional numbers in between that support the entire edifice. The relationist may say the question is irrelevant and arises from a misunderstanding – the series is simply a list of distinct entities as is demonstrated if the labelling is altered to a, b, c..

[§] For a good discussion of the historical development of the current world-view see Berkson[4].

For this reason it is still worthwhile considering if there are alternative ways of presenting Special Relativity, both to make it easier to apply and to identify faulty reasoning that leads to the apparent paradoxes.

### 3. Reference Frame

A reference frame is a standard against which motion is measured. A coordinate system is imposed on the frame to enable events and particle parameters to be quantified. The reference frame may adopt a variety of coordinate systems, but it is usual to describe space using three orthogonal Cartesian coordinates and assign a single coordinate to time. Events are then represented by 4-vectors. The *world* is the collection of all such events. However, it is not convenient for the human brain to manage the entirety of spacetime. Instead spacetime is foliated (or sliced) into a collection of spacelike hypersurfaces. All events on this hypersurface are described as simultaneous (the surface is also referred to as the hypersurface of simultaneity). This is most easily visualised using a space-time diagram. With the observer located at position $x = 0$ with proper time $t = t$, the hypersurface of simultaneity is represented by the horizontal line (red) in *Figure 1*. The three space dimensions are reduced to one for convenience.

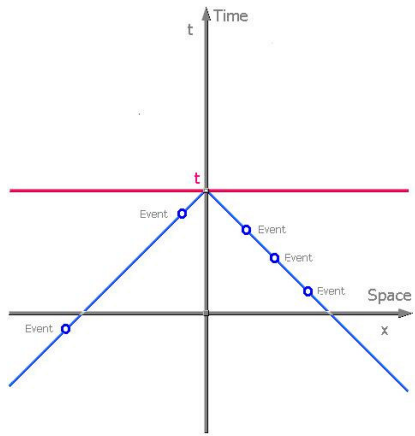

**Figure 1**  World map and world view

The set of events on the surface of simultaneity is known as the observer's *world map*. The hypersurface of simultaneity is constructed by imposing the proper time of the observer on all of space even though local time may possibly advance at a different rate at other locations. We recognise it as the normal surface to which the equations of physics are applied. For example, the definition of proper velocity as the change in 'proper' distance with respect to proper time is illustrated in *Figure 2*.

The use of the world map is based on an implicit assumption that the observer can travel freely at infinite speed. The frame cannot be maintained by measurement, only by the use of counterfactual statements and deduction. For example, one might say, 'I see that galaxy as it was one million years ago. If light travelled at an infinite speed, I would see that the galaxy is now much further away' and 'To find the real length of a rapidly moving object, I have to extrapolate the position of each end at the same time'.

(Lowe [10] examines in some detail the link between counterfactual conditionals and causality)

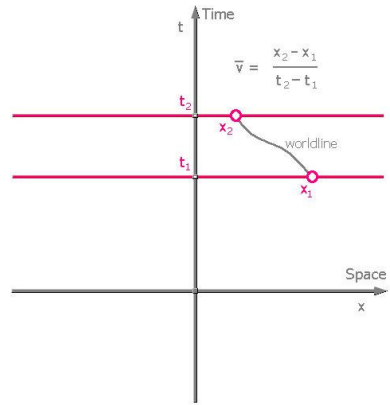

**Figure 2**  Definition of velocity

The world map has a unique and interesting property – it is the only way to slice spacetime such that every event on the surface is causally disconnected from all other events on the surface. It cannot be considered surprising therefore that the issue of the 'relativity of simultaneity' arises, where events that are simultaneous to a stationary observer may not be simultaneous a moving observer. After all, it is surely causal links that are significant as far as interaction is concerned and these are specifically excluded from the world map.

It is actually desirable to foliate spacetime in an alternative way to eliminate the artificial concept of simultaneity and thus remove many of the conceptual difficulties associated with Special Relativity. The interrelationship between events on the new surface should be unaffected by a change in inertial frame.

Null geodesics are events that are causally connected to the observer[§]. The hypersurface of null geodesics is represented by the two angled lines (blue) in *Figure 1*. All events on this line are causally connected to the observer and to one another (in the direction of positive time only). This spacetime slice is called the *world view*. It is proposed that this is a better way to slice spacetime because the property of connectedness that all events in the world view share is largely maintained following a velocity transformation. In simple terms, a world map does not transform to a world map (transformations of inertial frames will preserve the velocity of light but will not preserve simultaneity), but a world view does transform to a world view (over a restricted domain and range). Issues of measurement and how the same time can be maintained over the hypersurface no longer arise.

All events in the past and future that are causally connected to the observer will be referred to as the *causal event set*. The world view is a subset of the causal event set in that it only includes retarded events. We will calculate how the causal event set is affected by a velocity transform and solve some standard problems (including the twins problem) using this formalism. Of course, the results will be the same as derived in any standard Special Relativity textbook, e.g. Muirhead [11]; neither reference frames nor coordinate system have been altered and no changes have been made to the basic postulates.

---

[§] We will consider only the electromagnetic interaction in empty space which always propagates at velocity c. Causal contact thus only refers to events that are relateded in this way.

## 4. Interaction Continuity

As will be seen later, the issue of continuity becomes important when a causal event set is Lorentz transformed. The question here is whether interacting systems maintain continuity of contact through an acceleration. Consider the following scenario: An observer A is at rest with respect to entity B, a distance $x$ away. The observer receives a steady stream of photons from events initiated by B. The photons are received at times $t_i$ and were emitted at times $t_i - x/c$ from B's location. Contrast the situation where the observer is instantaneously accelerated at $t = 0$ and the situation where no acceleration takes place (*Figure 3*). We can imagine the photons that were incident upon the unaccelerated observer at $t = 0$ and absorbed at $t = 0_+$, an infinitesimal time afterwards. What will happen to these same photon(s) in the accelerated case?

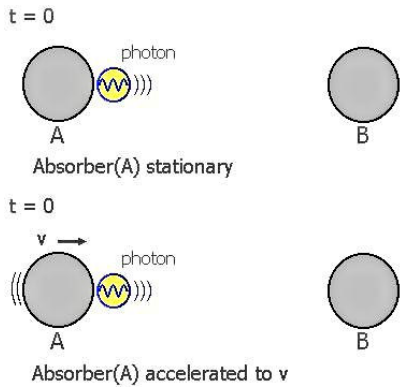

**Figure 3** Photon incident on accelerated absorber

The obvious response is that the photon is still absorbed, and indeed we will explore this assumption fully through the rest of the paper. The implication then is that interacting systems are continuous and causal contact is maintained irrespective of acceleration. However there are three other possibilities that do not violate energy conservation:

(A) **The photon passes through the absorber**. This would be the case if an accelerated entity experiences a time discontinuity with respect to B. For a short interval, A is not is causal contact. The photon must still move for this time and will thus move through the location of the absorber as if it were not there. Effectively the absorber disappears[¶]. This is an unlikely possibility, except that quantum mechanically particles are known to pass through apparently impenetrable barriers, the explanation in that case being the borrowing of energy by the Uncertainty Principle. The situation here is different and time discontinuities would surely have been noted as a deviation from the predictions of relativity in experiments measuring the lifetime of accelerated muons. Many materials are transparent to photons but surely none so selectively as implied here.

(B) **The photon is reflected or scattered by the absorber**. Does an accelerated entity act as a mirror? This again seems implausible. Classically, reflection takes place at a metallic surface because an electric field cannot exist within a metal – the mobile electrons in the free electron gas reconfigure to oppose the field. This imposes a boundary condition at or below the surface that gives rise to reflection. There is no evidence that reflection (or scattering) is associated with acceleration.

(C) **Photons that would have arrived at that time are not emitted.** A rather obvious explanation for quantum non-locality (the spooky collusion of paired photons over large distances) is to state that photon exchange is a 'prearranged' transaction between emitters and absorbers. Photons are emitted only because the absorbers will be at the right place at the right time with the correct properties. This global rather than local view of the electromagnetic interaction is totally deterministic but gives rise to even more difficult questions than are solved [¶]. For a transactional interpretation of this type, the notion of an emitted photon wandering through space until it chances upon an absorber is considered incorrect. The transactional interpretation is consistent with a relationist view of space. Curiously the world view that is being explored here eliminates the need for a propagation medium as the total mass-energy on each surface is constant over time. That is not the case with the world map – a photon emitted at $t = 0$ to be absorbed at $t = t$ will leave a mass-energy discrepancy at the times in between.

Of all the possibilities, the most plausible is that continuity is maintained. This is equivalent to the statement that the velocity of light is measured as constant in an accelerated frame. Whilst this assumption is taken forward, it should be born in mind that it may be incorrect – the assumption that the effects of acceleration is instantaneous must surely be considered further to determine how the effect of a continuously applied force can be partitioned into discrete instantaneous accelerations.

## 5. Definition of the Causal Event Set

The causal set of events is defined as the set of 4-vectors of all events causally connected to the observer. The relationist view of interaction may be incorporated by stating that events can only be associated with particles. The event set can be generalised to consider all potential events by including all particles at the particular moment of causal contact with the observer.

The analysis can be developed in one space dimension without loss of generality. Consider initially the *stationary frame* (the observer rest frame). The observer may arbitrarily be assigned a fixed space coordinate of $x = 0$. The proper time may be reset to $t = 0$ in the observer frame. All points in the frame share a common proper time and are stationary w.r.t. the observer. In general, each particle will contribute twice to the causal event set, once as an emitter and once as an absorber (retarded and advanced interactions, past and future). The causal event set may include reflection paths. If the condition is imposed that each slice will contain at least one transaction, it is clear that spacetime is sliced into a finite number of surfaces and that the foliation is not continuous. Of course it may be argued that surfaces with no energy transfer activity should still be included because particle positions may be altering hence there is a distinction between slices.

---

[¶] The absorber is effectively destroyed then recreated at a later time, an uncomfortable concept to associate with all accelerations.

[¶] For example, why can the probability function for spontaneous emission be correctly derived from the local Hamiltonian without reference to the distribution of distant absorbers?

If individual particles within the set of all observable particles $\wp$ are labelled $i$ with $x_i$ as the corresponding position relative to the observer (who is identified by $i = 0$) in the stationary frame, the retarded causal event set at proper time $t = 0$ is

$$S(0) = \{(x_i, -\frac{|x_i|}{c}) : i \in \wp, i \neq 0\} \quad (1)$$

Light signals from all these points will be received by the observer at $t = 0$. Position and time are no longer independent variables; one degree of freedom has been removed.

At a later time $t = t$, the retarded causal set is

$$S(t) = \{(x_i, t - \frac{|x_i|}{c}) : i \in \wp, i \neq 0\} \quad (2)$$

This is shown in *Figure 4*.

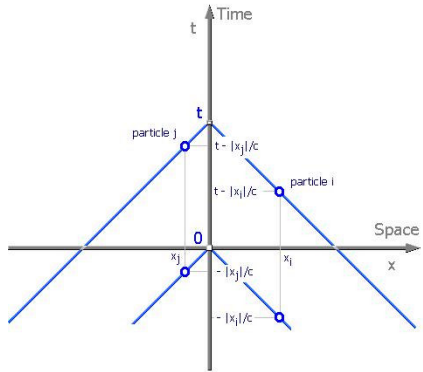

**Figure 4** Two members of the causal event set, events associated with particles *i* and *j*

### 6. The Lorentz Transform

Let the observer be accelerated at $t=0$ to a velocity $v$ instantaneously. This defines the *moving frame*. We will make the following assumption (as discussed in Section 4):

> For any particle in causal contact with the observer at a time prior to an acceleration, a signal transmitted an arbitrarily small time after the acceleration can also be received by the accelerated observer.

Before applying the transformations, the particles will be identified as being within one of two regions, $x_i \geq 0$ and $x_i < 0$, to eliminate the modulus bars and make clear the difference between the forward and reverse directions after acceleration.

At $t = 0$, members of the set of causally linked events transform as follows ($v > 0$) [11]:

$x_i \geq 0$

$$(x_i, -x_i/c) \to (\gamma[1+\beta]x_i, -\gamma[1+\beta]x_i/c) \quad (3)$$

$x_i < 0$

$$(x_i, x_i/c) \to (\gamma[1-\beta]x_i, \gamma[1-\beta]x_i/c) \quad (4)$$

Note that the new set of vectors also form a causal set therefore continuity is guaranteed. $\beta$ equals $v/c$ and $\gamma$ is $(1-\beta^2)^{-1/2}$. We can generalise by considering the transformation at a time $t$ following acceleration.

$x_i \geq 0$

$$(x_i, t - x_i/c) \quad (5)$$
$$\to (\gamma[1+\beta]x_i - \gamma vt, -\gamma[1+\beta]x_i/c + \gamma t)$$

$x_i < 0$

$$(x_i, t + x_i/c) \quad (6)$$
$$\to (\gamma[1-\beta]x_i - \gamma vt, \gamma[1-\beta]x_i/c + \gamma t)$$

We can show how events on the moving and stationary frames are dynamically related. The zero position in the moving frame maps to a specific point on a stationary causal set. Using the prime notation for the moving frame, $x' = 0$ requires that ($x_i \geq 0$)

$$\gamma(1+\beta)x_i = \gamma vt \quad (7)$$

or

$$x_i = \frac{vt}{1+\beta} \quad (8)$$

The event in the moving frame is therefore

$$E_R = (0, \gamma(1-\beta)t) \quad (9)$$

The connected event in the stationary frame is

$$E_R = (\frac{vt}{1+\beta}, \frac{t}{1+\beta}) \quad (10)$$

It is possible to define a retarded velocity, $v_R$, as the change in retarded position with proper time. From equation (8), calculating $|\beta|$ using the magnitude of $v$, the retarded speed when the object is moving away is

$$v_R = \frac{v}{1+|\beta|} \quad (11)$$

If the distant object is moving towards the observer the retarded velocity is

$$v_R = \frac{v}{1-|\beta|} \quad (12)$$

The diagram below summarises the relationship between the rest and moving causal event sets at acceleration point $t=0$ in the rest frame and at a later time $t = t$ measured in the rest frame (*Figure 5*).

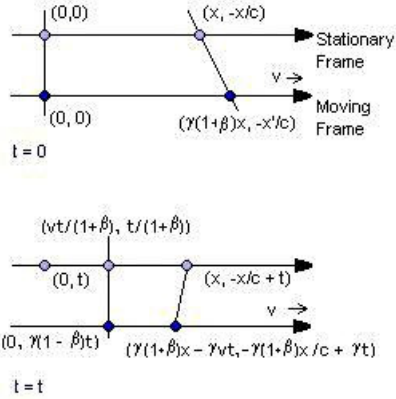

**Figure 5** The relationship between events in the rest and moving frames

It is important to recognise that the events sets for each observer are not the same (as demonstrated in section 8). The advantage of the use of the causal set is that a connected path of light signals can be defined which can pass between frames and correspond to real world measurements. We will show now how this can be applied to typical problems.

**7. The Twin 'Paradox'**

Consider again the scenario where twins A and B are initially at rest on earth. Let there be a beacon at the distance star C, the ultimate destination of twin B. The beacon is sending out a stream of photons which are perceived by both twins to originate from a source at distance $x$. Following the instantaneous acceleration of twin B, the top half of *Figure 5* (or equation (3)) shows that light from the beacon is now perceived to come from position $\gamma(1+\beta)x$. Using equation (12), the time taken to reach the beacon is found by dividing the retarded position by the retarded velocity (remembering that the beacon is moving towards twin B):

$$t_B = \frac{\gamma(1+\beta)x}{v/(1-\beta)} = \frac{x}{\gamma v} \qquad (13)$$

On reaching the beacon, twin B immediately decelerates to rest. The position of twin A is instantaneously transformed to $-x$. The time taken for twin B to reach the beacon from the viewpoint of twin A is simply

$$t_A = \frac{x}{v} \qquad (14)$$

There is clearly a difference in the elapsed time and there is no contradiction. The paradox disappears. The same argument can be used to show that the return journey contributes a similar difference.

Why then is the situation asymmetrical? It is because the spacetime of the accelerated twin is instantaneously transformed by the acceleration.

**8. Redshift**

The redshift is easy to understand with reference to the causal event set. No calculation is required. Looking at the bottom half of *Figure 5*, an interval $\gamma(1-\beta)t$ in the moving frame is causally linked to $t$ in the stationary frame when the object is moving away - the event $(0, \gamma(1-\beta)t)$ in the rest frame of the moving object is perceived to originate from the bridge point $(vt/(1+\beta), t/(1+\beta))$ by the observer. This is a member of the observer's causal event set at $t = t$. Thus for time t in the observer frame, a lesser time $\gamma(1-\beta)t = t/(1 + z)$ has elapsed in the moving frame, where $1+ z = (1+\beta)^{1/2}.(1-\beta)^{-1/2}$. If the frequency in the moving frame is $f$ then the frequency will register as $f/(1 + z)$ in the stationary frame. The wavelength increases hence we have a red shift for objects moving away.

Of course, the reciprocal must also be true: the moving observer must see light from the stationary observer as being red shifted too. This is easily demonstrated. Again from *Figure 5*, the bridge point between the frames is $(vt/(1+\beta), t/(1+\beta))$. Photons received by the moving observer will be part of the causal event set of the bridge point NOT of the stationary observer. (This change is key to the asymmetry). The stationary observer event that is part of the bridge causal set is

$$E_s = (0, \frac{t}{1+\beta} - \frac{vt}{(1+\beta)c}) \qquad (15)$$

or

$$E_s = (0, \frac{(1-\beta)t}{1+\beta}) \qquad (16)$$

which can be rewritten

$$E_s = (0, [\gamma(1-\beta)]^2 t) \qquad (17)$$

A time $t' = \gamma(1-\beta)t$ in the moving frame is associated with a time $\gamma(1-\beta)t'$ in the stationary frame. The accelerated observer will also measure an identical redshift. The situation is perfectly symmetrical and there is no way of determining which was accelerated after acceleration is over.

It is straightforward to demonstrate that photons received after acceleration are immediately wavelength-shifted. Consider three stationary spacecraft A, B and C aligned along an x-axis. A sends a continuous stream of photons with energy $E$ to B who immediately relays them on to C with a transponder. If B is suddenly accelerated to a velocity $v$ away from A, the received energy immediately changes to $E/(1 + z)$. This energy is passed onto C for whom the photon is blue-shifted by a factor $(1+z)$. The received energy is therefore $E$ as expected, the same as if spacecraft B were not there. In fact there is no way that a photon received directly from A can be distinguished from one relayed via a moving B could be distinguished by spacecraft C, even though they were emitted with different energies.

**9. Discussion**

By referring to *Figure 5*, it has been shown that two of the standard applications of Special Relativity can be solved trivially using the causal event set and with results that are totally consistent with the normal treatment. There are however conceptual issues arising that need to be examined further.

In essence, light signals that before an acceleration had originated from position $x$ (positive) and emitted at time $-x/c$ appear, after acceleration at $t = 0$, to have originated from position $\gamma(1+\beta)x$ at time $-\gamma(1+\beta)x/c$. For an accelerated object, the entire spacetime is instantly reconfigured. Following acceleration, the retarded 'position' of all previously stationary particles changes, distances being expanded in the forward direction and contracted in the reverse direction. This is a global and instantaneous change. Although interaction continuity is guaranteed during the acceleration, a space discontinuity results. The instantaneous nature of the transformation is required if the particles are not to have a 'memory' of accelerations and we can actually identify the principle of relativity with the fact that simple particles cannot maintain a memory of acceleration.

The second important point is that received photons are redshifted immediately following acceleration. Without absolute space or absolute acceleration, it is difficult to see how this can be compatible with a substantivalist view of spacetime.

The flexibility with which the space-time manifold is globally transformed would suggest that a relationist interpretation of space-time might be more appropriate than the substantivalist interpretation that was assumed at the beginning of the analysis. Because a photon cannot convey information about the source as it propagates, we are forced to seriously reconsider the transactional mechanism described earlier – that is to say, a photon transfer involves a handshake between the emitter and receiver and does not depend on a propagation medium. Again this suggests a relationist interpretation is correct. In this model, the speed of light is not something that arises from the movement of energy in spacetime but is the parameter from which spacetime is constructed by each observer to maintain the causal links between particles through an acceleration.

This is not really a comfortable conclusion because of the new problems that arise. In addition to the points made in Section 4(C), there is the issue of the loss of information. There is no doubt that the typical interaction process entails a high level of complexity, complexity that is commonly incorporated into the field. If space is removed, where does the complexity go? It is surely not feasible that elementary particles be conferred complex decision-making qualities¶. The only serious alternative is to assign the complexity to the Universe as a whole and attribute interaction to a holistic global mechanism in a universe consisting of a unity of particles in continuous causal contact. 'Fields' on the world map are then the projection of four-dimensional interactions onto a three dimensional hypersurface.

A simple analysis can be performed to calculate how the retarded position of events changes with time for a body subject to acceleration. If we consider an event at retarded position $x_R$ with an acceleration, $a$, in the direction of $x_R$, the apparent change in the event position with time is ($v$ positive)

$$\frac{dx_R}{dt} = \frac{dx_R}{dv} a - \frac{v}{(1-\frac{v}{c})} \quad (18)$$

Differentiating $x_R(v) = x_R(0)\gamma(1+v/c)$, and substituting

$$\frac{dx_R}{dt} = \frac{1}{1-\frac{v}{c}} \left( \frac{\gamma a x_R}{c} - v \right) \quad (19)$$

For $v/c \ll 1$, if we consider a Universe where all velocity is a result of the Hubble expansion ($v = H_o x \approx x/T$, where $H_o$ is the Hubble constant and $T$ is the current age of the universe) then the following choice of acceleration transforms the expanding Universe into a static Universe:

$$a = \frac{c}{T} = cH_o \quad (20)$$

Substituting equation (20) into equation (19) shows this to be true in the forward direction. It is easy to verify that this is true for all directions. The acceleration can thus be applied in any direction to effect the transformation.

This is the same acceleration that is considered significant in Milgrom's alternative description of the gravitational force, MOND. Sanders and McGaugh in their review of MOND[12] deem $cH_o$ to be a '*cosmologically interesting value*' of acceleration.

## 10. Mach's Principle

Does any of this have a bearing on the origin of inertia? Mach's principle states that in the absence of absolute (or any) space as a reference, the origin of inertia is the distant stars [13]. One aspect of inertia is the resistance of a body to a change in motion. There have been many attempts to quantify the principle by considering the effect of motion on distant matter (for example by Sciama[14]), all of which predict the right magnitude of response. The unsolved problem is that the inertial reaction force acts instantaneously. How is the effect of movement on distant objects going to work its way back on the instant the force is applied to generate an equal and opposite reaction force? One suggestion is the use of advanced and retarded waves, a concept introduced by Wheeler and Feynman[15] to explain electromagnetic radiation reaction, but this adds little to the transactional interpretation of energy transfer discussed earlier.

If for the moment we accept an instantaneous reconfiguration does take place with acceleration; is this not suggestive of inertia? It is an attractive idea, but an obvious mechanism for the reaction force does not emerge naturally. The distant matter is displaced but only in the frame of the accelerated observer – the actual matter will only be aware of the acceleration in the distant future. Certainly the reconfiguration of spacetime will alter the gravitational potential. To get a qualitative overview of the effect this might have, consider a Universe of constant density comprising particles at rest. If the observer is boosted to a velocity $v$, a mass element originally at distance $x$ in the forward direction moves to position $\gamma(1+\beta)x$. A mass element originally the same distance away in the reverse direction moves to $\gamma(1-\beta)x$. The distance between the two points has increased by a factor $\gamma$. However, this does not mean that the size of the observable universe has increased because, in reality, matter in the universe is not stationary. Objects at the boundary of the Universe are already moving a speed close to $c$ because of the Hubble expansion of the Universe: a small velocity boost superimposed in a recession velocity close to $c$ has little effect. Assuming that the mass elements are not significantly

---

¶ The search for an explanation for the workings of the Universe is really a process of removing from fundamental objects decision-making capabilities.

affected by the Hubble expansion, the average binding energy has increased:

$$E_G(v) = E_G(0) \frac{1}{1-\beta^2} \qquad (21)$$

If the total gravitational binding energy of the observer (mass m) with respect to the rest of the Universe (mass M) is given as

$$E_{TOTAL} = \frac{GMm}{\bar{r}} \qquad (22)$$

where r bar is some equivalent mean distance of the order of the radius of the universe, the total increase in gravitational binding energy§ to a second order approximation is

$$E_{FREE} = \frac{GMm}{\bar{r}} \cdot \frac{v^2}{c^2} \qquad (23)$$

This is of the same magnitude as the kinetic energy gain of the accelerated observer if we adopt Sciama's assumption[14] that the gravitational potential ($GM/x$) divided by $c^2$ is of the order one (certainly true if the current estimates for the baryonic mass and radius of the universe are slotted into the equation). The result though is unsatisfactory because $GM/x$ must not alter with time to conform with the Copernican Cosmological Principle (the Universe is homogeneous and isotropic), although one could always side with an anthropic ontology. There is no evidence that $G$ is changing with time[16], but $\bar{r}$ bar can remain constant within an expanding universe so long as mass is aggregating at short range to compensate for the outward movement of distant mass with time. This is then consistent with the cosmological principle, but is rather contrived and massive collaborative organisation on the part of the Universe is required to keep $\bar{r}$ bar constant.

If, in spite of these reservations, we can accept the calculation for the gravitational energy, it is straightforward to show that all energy may be gravitational in origin. Consider an inelastic collision of a mass $m_1$ moving at velocity $v_1$ and a stationary mass $m_2$. After collision, the first mass comes to a halt and the second mass moves off at a velocity $v_2 = m_1 v_1 / m_2$ (conservation of momentum). From equation (23), with $GM/rc^2$ represented by $k$, the total energy liberated from the gravitational potential is

$$\begin{aligned} E_{RELEASED} &= k(m_2 v_2^2 - m_1 v_1^2) \\ &= km_1 v_1^2 (\frac{m_1}{m_2} - 1) \end{aligned} \qquad (24)$$

exactly as expected in such a collision (if $k = \frac{1}{2}$). Masses appear to extract kinetic energy from their own frames through the gravitational potential.

The drawback of explaining inertia purely as the instantaneous redistribution of matter in the frame of the accelerated object is that the transformation effect is independent of mass. For this reason, Sciama's contention that inertia is related to the gravitational force, as was explored above, is more likely to be correct.

---

§ In this approximation, any energy associated with a particle moving with respect to a gravitational source has been ignored, but the relativistic mass increase is taken into account.